# Polaron Exchange Model for Ferromagnetic Ordering in Manganite Films

Liping Chen, Yubin Ma, Xianfeng Song, Guijun Lian, Guangcheng Xiong[a]

Department of Physics, Peking University, Beijing 100871, P. R. China

In doped manganites, the strong electron-phonon coupling due to the Jahn-Teller effect localizes the conduction-band electrons as polarons. This results in polarons are carriers responsible for transport and ferromagnetic ordering rather than the bare $e_g$ electrons, and sequentially polaron exchange model is emerged for describing ferromagnetic ordering. In $Pr_{0.7}(Sr_{1-x}Ca_x)_{0.3}MnO_3$(x=0.3-0.6) epitaxial thin films, for higher-temperature paramagnetic state and lower-temperature ferromagnetic state, both the temperature dependent transports present behaviors of small polaron; for paramagnetic-ferromagnetic transition, the experimental data of Curie temperature are well described by an energy balance expression induced by polaron exchange model. These results demonstrate that the polaron models are proper ways to describe the strongly correlated electrons in the doped manganites.

a) To whom correspondence should be addressed; e-mail: ssxgc@pku.edu.cn

The manganites $R_{1-x}A_x$MnO$_3$ have attracted considerable research interests in these years. *R* is a lanthanon element, and *A* is a divalent substituent such as Ca, Sr, Ba. At the doping level of x=0.3, these compounds present a paramagnetic-ferromagnetic transition accompanied by an insulator (d$\rho$/d$T$<0) to metal (I-M) (d$\rho$/d$T$>0) transition in cooling [1-8]. The I-M transition strongly depends on the applied magnetic field, thus leading to a large colossal magnetoresistance (CMR) effect near the transition temperature.

It is well accepted that the double exchange (DE) effect [9-11] in which $e_g$ electrons hop between Mn$^{3+}$ and Mn$^{4+}$ visa oxygen ions with a strong on-site Hund's-rule coupling [7, 8] is essential for understanding the magnetic and transport properties in manganites. In theoretical analysis, the DE exchange Hamiltonian item is defined as $H_{DE} = -t_{eff} \sum_{ij\sigma} (c_{i\sigma}^+ c_{j\sigma} + H.c) - J_H \sum_{i\alpha\beta} c_{i\alpha}^+ \sigma_{\alpha\beta} c_{j\beta} \cdot S_i$, where $J_H$ relates to Hund's rule coupling. Anderson and Hasegawa suggested that the strength of coupling for the FM ordering should depend on exchange matrix of the Hamiltonian which relies on the effective hopping of $e_g$ electrons [10]. However, when the DE effect with Hund's-rule coupling is solely considered, calculated resistivity is discrepant with the experimental data [12, 13]. Then, a very strong electron-phonon coupling coming from Jahn-Teller (J-T) lattice distortion was included [1].

In theoretical study, an additional item of J-T effect was introduced into Hamiltonian for understanding the conductivity and magnetoresistance effect [5], such as $H_{eff} = H_{DE}+H_{J-T}$. However, experimental data always exhibit complex pictures in which all kinds various interactions are merged. For example, changing doping level which mainly varies carrier density, bandwidth and Mn$^{3+}$/Mn$^{4+}$ ratio, is a valid method to affect the DE effect, but also brings change of the J-T lattice distortion. Therefore, in experiment, it is difficult to distinguish the contributions responding to different Hamiltonian items. In this case, simple models with few parameters may be expected for theoretical and experimental studies.

In strongly correlated electronic framework, for the existence of strong electron-phonon and other couplings, the transport carriers should not be bare $e_g$ electrons any more. The nature of the high-temperature phase of the manganites is controversial. The temperature dependences of the resistivity are generally discussed belonging to three models, i.e., a band-gap model, a variable-range hopping model, and a nearest-neighbor hopping model for the transport of small polarons. Some experimental support can be found in the literature for each model [14-16]. Since the properties of the manganites strongly depend on the microstructure and the oxygen content, investigations on various thin films of different quality have been performed. However, more and more studies suggested the manganites of high quality show the behavior of the small polaron models [7-8, 19-24]. The resistivity temperature dependence of $\rho(T)$ in the PM phase follows expression of $\rho(T) = \rho_0 T \exp(\frac{E_{hop}}{k_B T})$, where $E_{hop}$ is the hopping energy of the small polaron. In the model, the polarons are treated as quasi-particles that should overcome a potential barrier of $E_{hop}$. Samples of different $\rho$-$T$ data simply belongs to various $E_{hop}$ and $\rho_0$, and the influences of all kinds' complicated couplings are considered to be included into variation of polaron parameters, which greatly simplifies the complex picture of doped manganites.

In the small polaron model, most work has been devoted to the small-polaron thermal activated region. Recently, it was pointed out that the electrical transport behavior at low-temperature FM state was consistent with small-polaron coherent motion [25, 26]. The resistivity below 100 K obeys a formula of $\rho(T) = \rho_0 + E_1 \omega_S \sinh^2(\eta \omega_S / 2k_B T)$, where $\omega_S$ is the average frequency of the softest optical mode and $E_1$ is a constant proportional to the effective mass of polarons.

As a result, due to the strong electron-phonon coupling [9-11], the conducting band electrons responsible for transport are localized as carriers of polarons rather than the bare electrons. Both conductance and FM ordering originate from the same strongly correlated electrons, and I-M transition and FM ordering present related [2]. Therefore, the FM ordering should also relate to the polarons. In this paper, for $Pr_{0.7}(Sr_{1-x}Ca_x)_{0.3}MnO_3$ (PSCMO) thin films, we present that small polarons for transport and polaron exchange for FM ordering should be proper descriptions of the strongly correlated electrons in the manganites.

Epitaxial thin films of PSCMO(x=0.3-0.6) were deposited on (100)-oriented $SrTiO_3$ substrates as described previously [18]. The measurements of temperature dependent resistivity were performed by a standard four-terminal method in different constant magnetic fields from 0T up to 8T in steps of 1T even smaller.

Figure 1(a) shows the $\rho(T)$ curves in zero field for PSCMO(x=0.3-0.6) films. The M-I transition temperature ($T_P$) decreases from 239 K for x = 0.3 to 227 K for x = 0.4 and then to lower one of 170 K for x =0.6. The Curie temperature ($T_C$) obtained from resistivity measurement defined as the maximum in $d(log\rho)/dT$ [3] is close to that obtained in magnetization measurement [19]. It gets smaller with the increasing x from 218 K for x = 0.3 to 145K for x =0.6. This result indicates that the contribution of the J-T lattice distortion on I-M transition and FM ordering is adjusted by varying the average ion radius of Pr site with adjusting the Ca substitution concentration of x.

Fig. 1(b) presents the $\rho(T)$ data of PSCMO (x = 0.4) film in magnetic fields. The enhanced magnetic field results in a shift of $T_P$ to higher temperature and a decreasing of film resistivity in transition area, showing a typical CMR effect. As displayed in the Fig.1(c), the linear $\ln(\rho/T)$-$1/T$ curves in the PM phases present well-fitted behavior for the adiabatic small polaron. Obtained from the slopes of lines, the hopping energy of the small polaron $E_{hop}$ is found to decrease with the increasing magnetic fields. In the low-temperature FM state, as examples, the $\rho(T)$ curves in 0 and 8T magnetic fields are magnified in the Fig.1(d). They are quite consistent with the fitting curves of small polaron coherent motion. Similar behaviors of small polarons, thermal activation in PM state and coherent motion in low-temperature FM state, are observed in transport of the other films. These indicate that, due to strong electron-phonon coupling, the carriers in PM state and low-temperature FM state for transport should be small polarons as discussed previously [7-8, 20-26].

Since strongly correlated $e_g$ electrons that present behaviors of small polarons in conductance are also related to FM ordering in the films. Polarons are suggested the carriers of FM ordering, and a polaron exchange model is sequentially emerged. In the model, the motion of the self-localized carrier is dominated by hopping process where the polaron as quasi-particle has to overcome a potential barrier of $\Delta E$. The kinetic energy caused by effective exchange of polarons for the FM ordering should be proportional to the exchange probability of $\exp(-\Delta E/k_B T)$. Once the energy for the FM ordering equals to the thermal activation energy of $k_B T$, FM ordering will get steadied. For that, there will be an energy balance expressed as

$k_B T_C = E_0 \exp(-\Delta E/k_B T_C)$.----------------------------------(1)

where $E_0$ is the energy for FM ordering provided by the effective polarons exchange, and the $\Delta E$ denotes the potential barrier to be overcome by the exchange polarons at $T_C$.

The inset of Fig. 2 shows the plot of the energy balance expression with variable $\Delta E$ for $y=k_BT_C$ and unchanged $E_0=1$. At beginning, $T_C$ decreases gently when $\Delta E$ increases. Then, with further increasing $\Delta E$, the energy balance condition results in a rapid drop of $T_C$, and leads to a threshold of $\Delta E$ at $\Delta E=k_BT_C$. At the threshold, the differential coefficient of $dT_C/d\Delta E$ becomes negative infinite. As a result, once the $\Delta E$ increases from this point, the $T_C$ will drop to zero. In other words, it indicates that if the potential barrier of $\Delta E$ is higher than the threshold value, the amount of effective exchanged carriers will be less than the required for keeping the FM ordering stable, and the FM transition should not be obtained.

It is interesting and wondering that the behaviors of $T_C(H)$ versus $E_{hop}(H)$ for the films are similar to that of $T_C$ versus $\Delta E$ in the energy balance expression. The experimental data $T_C$-$E_{hop}$ for a PSCMO (x = 0.4) thin film in magnetic fields is shown in Fig. 2. Same as the $T_C$ versus $\Delta E$ in the inset, in lower magnetic fields (0<$H$<1.5 T), $T_C(H)$ increases quickly with decreasing $E_{hop}(H)$; In further increasing magnetic fields (2<$H$<8.5 T), $T_C(H)$ becomes to increase slowly with further decreasing $E_{hop}(H)$. This behavior implies that the trend of $E_{hop}$ in magnetic fields is similar to $\Delta E$.

On the assumptions that in the applied relatively lower magnetic fields $E_0(H)$ changes little for a sample and $\Delta E(H)$ depends on $E_{hop}(H)$ as

$$\Delta E(H)/\Delta E(H=0)+\Delta E_0=(E_{hop}(H)/E_{hop}(H=0))/t \text{----------------------------------(2)}$$

where $\Delta E_0$ is a constant shift from the zero point, we used the energy balance expression to fit the experimental data of $T_C(H)$-$E_{hop}(H)$, in which parameters of the $E_0(H)$ and $\Delta E(H)$ were evaluated. Table I presents the detail data for the four PSCMO thin films in the applied magnetic fields. Fig.3 shows the experimental data and the fitting curves for the films. All the films' experimental data of $T_C(H)$-$E_{hop}(H)$ are quite consistent with the fitting curves. These results indicate that the energy balance expression together with the two assumptions is reasonable in physics and valuable in understanding the manganite films.

For a sample, the $E_0(H)$ calculated from the expression(1) with $\Delta E(H)$ is little changed in the applied fields, and almost equals to $E_0=ek_BT_C$ from the threshold point in the fitting curve in Fig.3. This is self-consistent with the assumption that $E_0(H)$ changes little in the applied fields for a sample. In different magnetic fields, various $T_C$ were obtained. However, the various $T_C$ for a sample fall into a merely changed $E_0$, which demonstrated that the $E_0$ change little for different temperature around the transition. For the PSCMO(x=0.3-0.6) films, obtained in Table I, The film with less doping level of Ca has a higher $E_0$. With a same exchange probability, the energy of the polaron exchange devoting to FM ordering is determined by $E_0$. The larger $E_0$ is, the stronger the coupling of FM ordering is. The $E_0$ reflects the intrinsic strength of FM coupling for a material. Enhanced average $R$-site radius $<r_R>$ due to the decreasing Ca doping level induces a wider Mn-O-Mn angle and longer Mn-O bond distance. Therefore, in micromechanics, $E_0$ may be related to the structure of lattice such as angle of Mn-O-Mn and Mn-O bond distance.

The $\Delta E(H)$ at $T_C$ present dependent on $E_{hop}(H)$, this supported assumption demonstrates that FM ordering caused by polaron exchange relates with transport properties. As a potential barrier for the hopping of small polaron, the decreasing $E_{hop}$ in an magnetic field indicates a drop of the potential barrier, which results in easier hopping for the small polarons, and sequentially a drop of the resistivity. Similarly, a barrier potential exists in the polaron exchange, and the magnetic field also reduces the barrier, so the film in magnetic field will get into FM state at a higher temperature. Though, the $E_{hop}$ and $\Delta E$ are responsible for transport and FM ordering respectively, their similar behaviors in the magnetic fields indicated that both the polarons for transport and FM ordering originate same from the strong correlated electrons.

All the experimental data exhibit quite similar characters that $T_C(H)$ slowly decreases with increasing $\Delta E(H)$, and quickly decreases with further increasing $\Delta E(H)$ in lower magnetic fields. Each of them is similar to the energy balance expression: FM ordering temperature of $T_C$ in 0T field shows the threshold condition of $\Delta E(0)=k_BT_C(0)$, and the $T_C$ increases from the threshold point, which displays that FM ordering collapses above the evaluated $\Delta E(0)$. The energy balance curve with threshold character suggests that, with decreasing potential barriers for transport and exchange, FM ordering is achieved when the energy balance condition is reached.

The phase separation together with the percolation conductivity is an intrinsic feature of doped manganites and it may even lie at the very core of the CMR phenomenon. It is noticeable that all our experimental data exhibit quite consistent characteristic with the polaron exchange model. Though there are coexistences of multiphase and conductive percolation in films. Therefore, it is demonstrated that the polaron models in which interactions for strongly correlated electron are integrated in polarons are able to reflect the effective Hamiltonian. The influences of magnetic fields on multiphase such as volume ratio of FM phase to PM phase are included in the $\Delta E(H)$ of polarons. As a result, the polarons exchange model with the energy balance expression would be a simple and quite appropriate approach for describing the FM ordering in the manganite films.

In a summary, for PSCMO thin films, temperature dependent resistivity of $\rho(T)$ exhibits the small polaron behaviors in high-temperature PM states and low-temperature FM states. On the other hand, we present a polaron exchange model for FM ordering, which is consistent with the experimental data. These results indicate that polarons should be a better description of the strongly correlated electrons for transport and FM ordering in manganites.

We thank the support of the National Natural Science Foundation of China.

**Figure Captions**

FIG. 1: The $\rho(T)$ curves in zero field for PSCMO(x=0.3-0.6) films (a); The $\rho(T)$ data measured in magnetic fields (b), the linear $\ln(\rho/T)$-$1/T$ curves in the *PM* phase(c), the $\rho(T)$ curves in low-temperature FM state and the fitting curve obtained from the expression of small polaron coherent motion (d) for PSCMO(x=0.4) film.

FIG. 2: The experimental data $T_C$-$E_{hop}$ for PSCMO (x=0.4) thin film in magnetic fields; the inset shows the $T_C$ versus $\Delta E$ in the energy balance expression.

FiIG. 3: The experimental data of $T_C(H)$-$E_{hop}(H)$ for the films and the fitting curves of the energy balance expression.

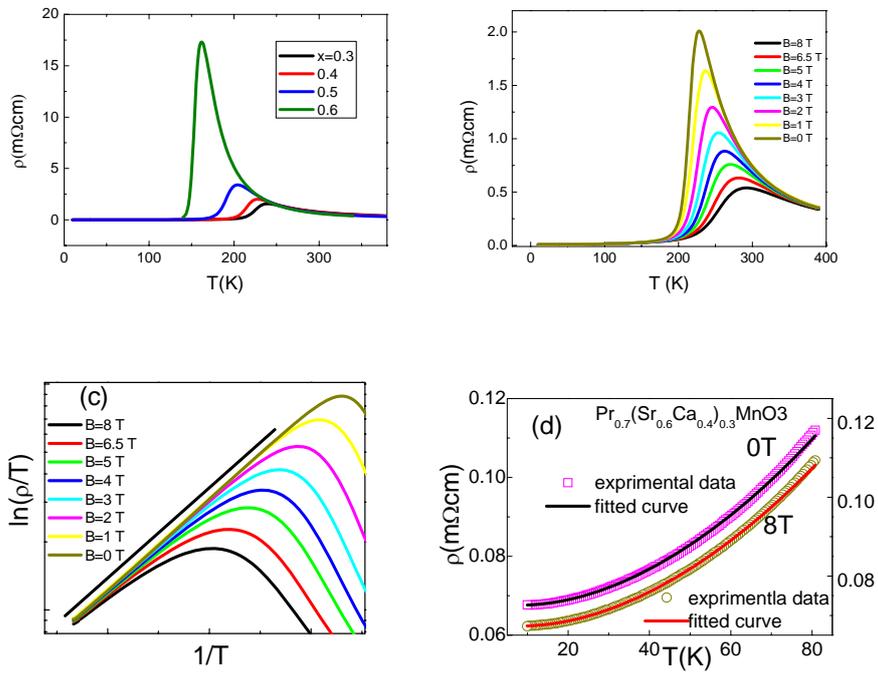

Figure 1

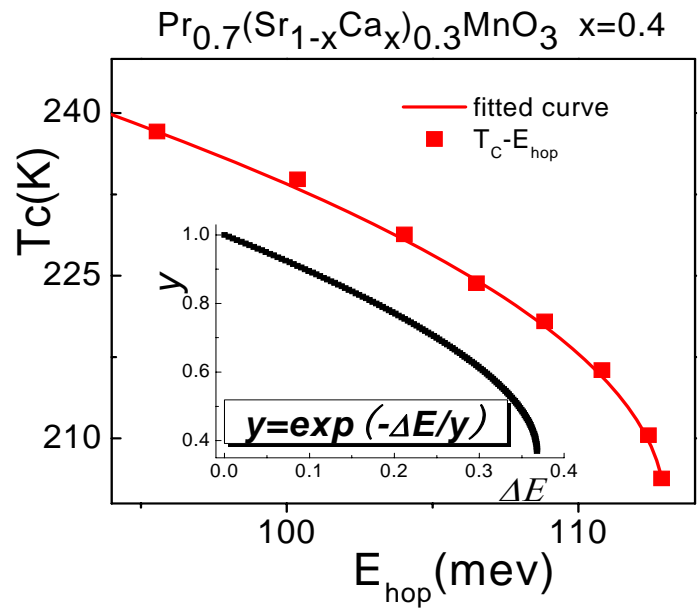

Figure 2

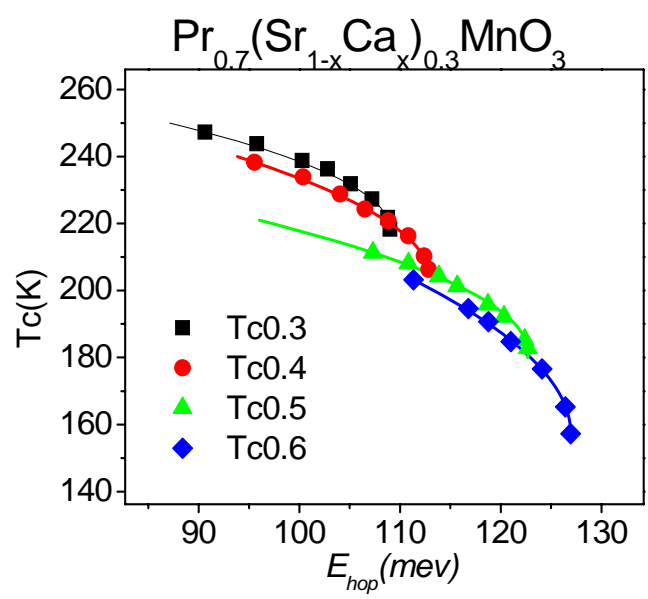

Figure 3

TABLE I. Experimental data of $T_C$ for the PSCMO thin films in 0 and 8 T magnetic fields and their relevant fitting parameters: hopping energy $E_{hop}$ for small polaron, $\Delta E$ and $E_0$ for polaron exchange.

|  | $T_C(H)$ (K) | | $E_{hop}(H)$ (meV) | | $\Delta E(H)$ (meV) | | $E_0(H)$ (meV) | |
| --- | --- | --- | --- | --- | --- | --- | --- | --- |
| $H$ | 0 T | 8 T | 0 T | 8 T | 0 T | 8 T | 0 T | 8 T |
| PSCMO (0.3) | 218.3 | 247.3 | 108.99 | 90.62 | 18.82 | 18.66 | 51.16 | 51.16 |
| PSCMO (0.4) | 206.3 | 238.3 | 112.85 | 95.56 | 17.57 | 17.33 | 47.76 | 47.76 |
| PSCMO (0.5) | 182.8 | 211.3 | 122.71 | 107.30 | 15.66 | 15.46 | 42.56 | 42.56 |
| PSCMO (0.6) | 145.7 | 187.3* | 132.16 | 116.29* | 12.70 | 12.27* | 34.52 | 34.52* |

\* The magnetic field of $H$ is 7 T.